\documentclass[aps,preprint]{revtex4}%
\usepackage{amsfonts}
\usepackage{amsmath}
\usepackage{amssymb}
\usepackage{graphicx}
\usepackage{epstopdf}%
\providecommand{\U}[1]{\protect\rule{.1in}{.1in}}
\begin{document}
\title{Damping control in viscoelastic beam dynamics}
\author{Elena Pierro$^{1}$}
\affiliation{$^{1}$Scuola di Ingegneria, Universit\`{a} degli Studi della Basilicata, 85100
Potenza, Italy}
\keywords{beam dynamics, viscoelasticity, modal analysis, damping identification,
damping control, linear systems}
\pacs{PACS number}

\begin{abstract}
Viscoelasticity plays a key role in many practical applications and in
different reasearch fields, such as in seals, sliding-rolling contacts and
crack propagation. In all these contexts, a proper knowledge of the
viscoelastic modulus is very important. However, the experimental
characterization of the frequency dependent modulus, carried out through
different standard procedures, still presents some complexities, then possible
alternative approaches are desirable. For example, the experimental
investigation of viscoelastic beam dynamics would be challenging, especially
for the intrinsic simplicity of this kind of test. This is why, a deep
understanding of damping mechanisms in viscoelastic beams results to be a
quite important task to better predict their dynamics. With the aim to
enlighten damping properties in such structures, an analytical study of the
transversal vibrations of a viscoelastic beam is presented in this paper. Some
dimensionless parameters are defined, depending on the material properties and
the beam geometry, which enable to shrewdly design the beam dynamics. In this
way, by properly tuning such disclosed parameters, for example the
dimensionless beam length or a chosen material, it is possible to enhance or
suppress some resonant peaks, one at a time or more simultaneously. This is a
remarkable possibility to efficiently control damping in these structures, and
the results presented in this paper may help in elucidating experimental
procedures for the characterization of viscoelastic materials.

\end{abstract}
\startpage{1}
\endpage{2}
\maketitle

\section{Introduction}

\begin{figure}[ptb]
\begin{center}
\includegraphics[
height=5cm,
]{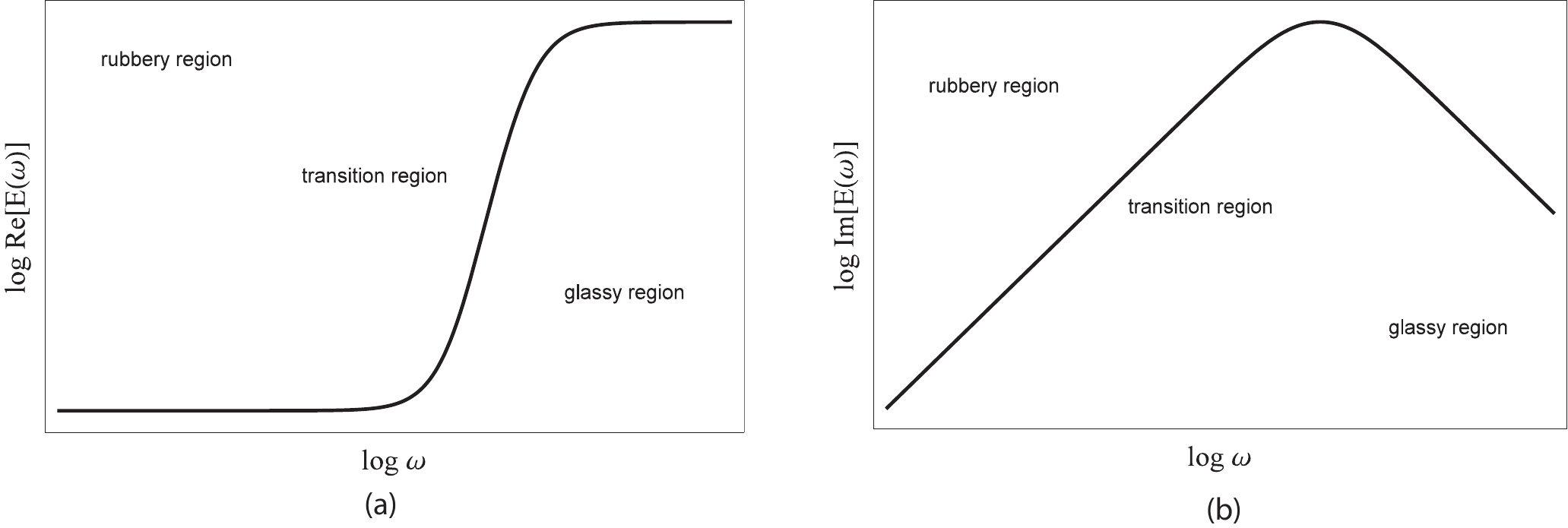}
\end{center}
\caption{The real part (a) and the imaginary part (b) of the elastic modulus
$E\left(  \omega\right)  $ of a generic viscoelastic material.}%
\label{Figure1}%
\end{figure}

Nowadays, viscoelastic materials are widely utilized in several engineering
applications, such as seals \cite{Bottiglione2009} and adhesives/biomimetic
adhesives \cite{Carbone2011,Carbone2012,Carbone2012bis,Carbone2013bis}.
Moreover, they are object of recent research investigations, for example: (i)
rolling contacts \cite{Persson2010,Dumitru2009,Carbone2013}, (ii) sliding
contacts \cite{Grosch1963,Carbone2004,Carbone2009,Persson2001}, (iii) crack
propagation \cite{Carbone2005,Carbone2005bis,Persson2005}, (iv) viscoelastic
dewetting transition \cite{Carbone2004bis}. In all the aforementioned research
fields, having knowledge of the correct viscoelastic modulus in the frequency
domain is of utmost importance. Very often, viscoelastic materials are also
combined with fibers or fillers, but also in this case, the mechanical
behaviour of the viscoelastic matrix must be well established, especially as
input data for numerical simulations. Usually, viscoelastic modulus is
experimentally characterized, and one of the most utilized technique is the
DMA (Dynamic mechanical analysis) \cite{Rasa2014}, which is quite complex and
time consuming. Alternative approches have been presented in literature, such
as the experimental dynamic evaluation of the viscoelastic beam-like
structures \cite{Caracciolo1,Caracciolo2,Cortes2007}. Such experiments,
however, require a good comprehension from a theoretical point of view of the
viscoelastic beam dynamics. Many theoretical studies on the dynamics of
non-viscous damped oscillators, for both SDOF
\cite{Muller2005,Barruetabena2011}, and MDOF \cite{Adhikari2002,Lazaro2013}
systems, have been presented in literature. Also viscoelastic continuous
systems have been theoretically and experimentally investigated in their
dynamics, such as beams and plates \cite{Barruetabena2012,Inman1989,Gupta2007}%
. However, most of these studies do not present a qualitative analysis of the
dynamic characteristics of such systems, in terms of eigenvalues and their
connection with the most representative physical parameters. Only in Ref.
\cite{Adhikari2005}, a deep analysis of a single degree-of-freedom
non-viscously damped oscillator has been presented. Extending this kind of
investigation to continuous systems would be of crucial concern when
viscoelastic properties of materials must be properly established. With the
aim to shed light on the vibrational behaviour of such systems, in this paper
a detailed study of the dynamics of a viscoelastic beam is presented. Recall
that the viscoelastic materials are characterized by the most general
stress-strain relation \cite{Christensen}
\begin{equation}
\sigma\left(  x,t\right)  =\int_{-\infty}^{t}G\left(  t-\tau\right)
\dot{\varepsilon}\left(  x,\tau\right)  \mathrm{d}\tau\label{stress-strain}%
\end{equation}
where $\dot{\varepsilon}(t)$ is time derivative of the strain, $\sigma(t)$ is
the stress, $G\left(  t-\tau\right)  $ is the time-dependent relaxation
function, which is related, in the Laplace domain, to the viscoelastic modulus
$E\left(  s\right)  $ through the relation $E\left(  s\right)  =sG\left(
s\right)  $. Usually, a discrete version of $E\left(  s\right)  $ is utilized
to characterize linear viscoelastic solids, which can be represented in the
Laplace domain as%
\begin{equation}
E\left(  s\right)  =E_{0}+\sum_{k}E_{k}\frac{s\tau_{k}}{1+s\tau_{k}%
}\label{ElasticModulusLaplace}%
\end{equation}
where $E_{0}$ is the elastic modulus of the material at zero-frequency,
$\tau_{k}$ and $E_{k}$ are the relaxation time and the elastic modulus
respectively of the generic spring-element in the generalized linear
viscoelastic model \cite{Christensen}. The general trend of the viscoelastic
modulus $E(\omega)$ is shown in Figure \ref{Figure1}. It can be observed that
at low frequencies the material is in the `rubbery' region, indeed
$\operatorname{Re}[E(\omega)]$ is relatively small and approximately constant
(Figure \ref{Figure1}-a), and the viscoelastic dissipations related to the
imaginary part $\operatorname{Im}[E(\omega)]$ of the viscoelastic modulus
becomes negligible (Figure \ref{Figure1}-b). At very high frequencies the
material is elastically very stiff (brittle-like). In this `glassy' region
$\operatorname{Re}[E(\omega)]$ is again nearly constant but much larger
(generally by 3 to 4 orders of magnitude) than in the rubbery region. The
intermediate frequency range (the so called `transition' region) determines
the energy dissipation, and can completely deviate the modal behaviour of a
viscoelastic solid from the equivalent elastic one. Moreover, the transition
region, and hence the functions $\operatorname{Re}[E(\omega)]$ and
$\operatorname{Im}[E(\omega)]$, can be shifted towards higher or smaller
frequencies by simply varying temperature, because of the viscoelastic modulus
$E(\omega)$ dependence on temperature \cite{Christensen}. Of course, only the knowledge of the analytical vibrational
response of a viscoelastic structure can provide the right parametric
quantities, useful to accurately enlighten the relationship between the
material properties and the modal contents. In this direction, the flexural
vibrations of a viscoelastic beam is analytically studied in this paper, and
by introducing some non-dimensional parameters, a qualitative analysis of the
eigenvalues is presented. At first, an ideal viscoelastic material is
considered, i.e. characterized by one single relaxation time. This kind of
study, indeed, is useful for a first understanding of the physical parameters
enclosed in the problem. Then, two relaxation times are taken into account,
and their influence on the dynamics of the beam is deeply evaluated and
described. Finally, some considerations are pointed out regarding the
vibrational response of the beam in case of real viscoelastic materials.

\section{The Model}

In this section the analytical dynamic response of a viscoelastic beam with
rectangular cross section is derived. Let be $L$, $W$, and $H$ respectively
the length, the width and the thickness of the beam (Figure \ref{Figure2}),
and let us assume that $L\gg W$, $L\gg H$. Assuming also that the displacement
along the $z$-axis $\left\vert u\left(  x,t\right)  \right\vert \ll L$, the
Bernoulli theory of transversal vibrations can be applied and therefore it is
possible to neglect the influence of shear stress in the beam. It is worth
noticing that this hypotesis does not limit the validity of the analysis,
since the attention is paid to the first resonant peaks, which are not
affected by shear deformations. Hence, the general equation of motion is
\cite{Inman1996}%
\begin{equation}
J_{xz}\int_{-\infty}^{t}E\left(  t-\tau\right)  \frac{\partial^{4}u\left(
x,\tau\right)  }{\partial x^{4}}\mathrm{d}\tau+\mu~\frac{\partial^{2}u\left(
x,t\right)  }{\partial t^{2}}=f\left(  x,t\right)  \label{motionEq}%
\end{equation}
where $\mu=\rho A$, $\rho$ is the bulk density of the material the cantilever
is made of, $A$ is the area of the cross section of the beam, i.e. $A=WH$,
$J_{xz}=(1/12)WH^{3}$, and $f\left(  x,t\right)  $ is the generic forcing
term. It must be highlighted that some additional terms could be considered in
Eq.(\ref{motionEq}), representing different kind of damping contributes
\cite{Banks91} (e.g. viscous damping and hysteresis damping). In the present
study such terms are neglected, but it is important to underline that the
results obtained in this paper are not affected by this assumption from a
qualitative point of view. The forced solution of the above problem
Eq.(\ref{motionEq}) can be found in the form of a series of the eigenfunction
$\phi_{n}\left(  x\right)  $ of the following problem
\begin{equation}
J_{xz}\int_{-\infty}^{t}E\left(  t-\tau\right)  u_{xxxx}\left(  x,\tau\right)
\mathrm{d}\tau+\mu~u_{tt}\left(  x,t\right)  =0 \label{autoproblem}%
\end{equation}
($u_{x}\left(  x,t\right)  =\partial u\left(  x,t\right)  /\partial x$,
$u_{t}\left(  x,t\right)  =\partial u\left(  x,t\right)  /\partial t$), with
the opportune boundary conditions. In this study, the free-free boundary
conditions are considered
\begin{align}
u_{xx}\left(  0,t\right)   &  =0\label{BCautoproblem}\\
u_{xxx}\left(  0,t\right)   &  =0\nonumber\\
u_{xx}\left(  L,t\right)   &  =0\nonumber\\
u_{xxx}\left(  L,t\right)   &  =0\nonumber
\end{align}
. By Laplace transforming the time-dependence in Eq.(\ref{autoproblem}), and
considering equal to zero the initial conditions, it is easy to show that the
eigenfunctions $\phi\left(  x,s\right)  $ must satisfy the following equation%
\begin{equation}
\phi_{xxxx}\left(  x\right)  -\beta_{eq}^{4}\left(  s\right)  \phi\left(
x\right)  =0 \label{autoproblemLaplace}%
\end{equation}
where it is defined%
\begin{equation}
\beta_{eq}^{4}\left(  s\right)  =-\frac{\mu~s^{2}}{J_{xz}E\left(  s\right)
}=-\frac{\mu~s^{2}}{J_{xz}}C\left(  s\right)  \label{beta_equivalent}%
\end{equation}
and the compliance of the viscoelastic material $C\left(  s\right)
=1/E\left(  s\right)  .$ The boundary conditions then become%
\begin{align}
\phi_{xx}\left(  0\right)   &  =0\label{BC autofunction}\\
\phi_{xxx}\left(  0\right)   &  =0\nonumber\\
\phi_{xx}\left(  L\right)   &  =0\nonumber\\
\phi_{xxx}\left(  L\right)   &  =0\nonumber
\end{align}

The solution of the above Eq.(\ref{autoproblemLaplace}) can be written in the
form%
\begin{equation}
\phi(x,s)=W_{1}\cos\left[  {\beta}_{eq}{\left(  s\right)  x}\right]
+W_{2}\sin\left[  {\beta}_{eq}{\left(  s\right)  x}\right]  +W_{3}\cosh\left[
{\beta}_{eq}{\left(  s\right)  x}\right]  +W_{4}\sinh\left[  {\beta}%
_{eq}{\left(  s\right)  x}\right]
\end{equation}
and by requiring that the determinant of the system matrix obtained from
Eqs.(\ref{BC autofunction}) is zero, one obtains%
\begin{equation}
\left[  1-\cos\left(  \beta_{eq}L\right)  \cosh\left(  \beta_{eq}L\right)
\right]  =0 \label{modes_eq}%
\end{equation}
The solutions $\beta_{n}L=c_{n}$ of the above Eq.(\ref{modes_eq}) are well
known \cite{Inman1996}, and they are the same of the perfectly elastic case.
In particular, from the following relation%
\begin{equation}
-\frac{\mu~s^{2}}{J_{xz}E\left(  s\right)  }=\left(  \beta_{n}\right)
^{4}=\left(  \frac{c_{n}}{L}\right)  ^{4} \label{beta_solutions}%
\end{equation}
it is possible to calculate the complex conjugate eigenvalues $s_{n}$
corresponding to the $n_{th}$ mode, and the real poles $s_{k}$ related to the
material viscoelasticity (a detailed analysis of the eigenvalues will be shown
in the next section). The values $\beta_{n}$ allow to determine the
eigenfunctions $\phi_{n}\left(  x\right)  $%
\begin{equation}
\phi_{n}\left(  x\right)  =\cosh\left(  \beta_{n}x\right)  +\cos\left(
\beta_{n}x\right)  -\frac{\cosh\left(  \beta_{n}L\right)  -\cos\left(
\beta_{n}L\right)  }{\sinh\left(  \beta_{n}L\right)  -\sin\left(  \beta
_{n}L\right)  }\left[  \sinh\left(  \beta_{n}x\right)  +\sin\left(  \beta
_{n}x\right)  \right]  \label{modes}%
\end{equation}
which are equal to the eigenfunctions of the elastic case. A simple proof of
the previous statement can be shown by considering the initial conditions
$u\left(  x,0\right)  =\phi_{n}\left(  x\right)  $ and $u_{t}\left(
x,0\right)  =0$ of the problem Eq.(\ref{autoproblem}). In this case, indeed,
the solution of Eq.(\ref{autoproblem}) is $u\left(  x,t\right)  =C~\mathrm{e}%
^{\operatorname{Re}\left[  s_{n}\right]  t}~\phi_{n}\left(  x\right)
~\cos\left(  \operatorname{Im}\left[  s_{n}\right]  t\right)  $.

\section{Beam response}

In this section the solution of Eq.(\ref{motionEq}) is calculated, by
considering (see Ref.\cite{Inman1989}) the decomposition of the system
response into the modes $\phi_{n}\left(  x\right)  $ of the beam%
\begin{equation}
u\left(  x,t\right)  =\sum_{n=1}^{+\infty}\phi_{n}\left(  x\right)
q_{n}\left(  t\right)  \label{sepvariable}%
\end{equation}
For the orthogonality condition one has%
\begin{equation}
\frac{1}{L}\int_{0}^{L}\phi_{n}\left(  x\right)  \phi_{m}\left(  x\right)
\mathrm{d}x=\delta_{nm} \label{orthogonality}%
\end{equation}
where $\delta_{nm}$ is Kronecker delta function. Moreover, because of
Eqs.(\ref{autoproblemLaplace})-(\ref{beta_equivalent}), the following relation
holds true
\begin{equation}
\frac{1}{L}\int_{0}^{L}\left(  \phi_{n}\right)  _{xxxx}\left(  x\right)
\phi_{m}\left(  x\right)  \mathrm{d}x=\frac{1}{L}\int_{0}^{L}\phi_{n}\left(
x\right)  \beta_{n}^{4}\phi_{m}\left(  x\right)  \mathrm{d}x=\delta_{nm}%
\beta_{n}^{4} \label{orthog_der}%
\end{equation}
Let us project the equation of motion on the function $\phi_{m}\left(
x\right)  $ of the basis. The projected solution $u_{m}\left(  t\right)  $ is
defined as%

\begin{equation}
u_{m}\left(  t\right)  =\left\langle u\left(  x,t\right)  \phi_{m}\left(
x\right)  \right\rangle =\frac{1}{L}\int_{0}^{L}u\left(  x,t\right)  \phi
_{m}\left(  x\right)  \mathrm{d}x \label{sol_projected}%
\end{equation}
therefore Eq.(\ref{motionEq}) becomes%
\begin{equation}
\mu\ddot{q}_{n}\left(  t\right)  +J_{xz}\beta_{n}^{4}\int_{-\infty}%
^{t}E\left(  t-\tau\right)  q_{n}\left(  \tau\right)  \mathrm{d}\tau
=f_{n}\left(  t\right)  \label{Eq projected time}%
\end{equation}
where $f_{n}\left(  t\right)  =\frac{1}{L}\int_{0}^{L}f\left(  x,t\right)
\phi_{n}\left(  x\right)  \mathrm{d}x$ is the projected force term. By taking
the Laplace Transform of Eq.(\ref{Eq projected time}), with initial conditions
equal to zero, one obtains%
\begin{equation}
\mu s^{2}Q_{n}\left(  s\right)  +J_{xz}\beta_{n}^{4}E\left(  s\right)
Q_{n}\left(  s\right)  =F_{n}\left(  s\right)  \label{Eq projected Laplace}%
\end{equation}
It is possible to rewrite the above equation as%
\begin{equation}
Q_{n}\left(  s\right)  =H_{n}\left(  s\right)  F_{n}\left(  s\right)
\label{Qn}%
\end{equation}
where the function%
\begin{equation}
H_{n}\left(  s\right)  =\frac{1}{\left[  \mu s^{2}+J_{xz}\beta_{n}^{4}E\left(
s\right)  \right]  } \label{transfer function}%
\end{equation}
is the Transfer Function of the system, for the $n_{th}$ mode.

Eq.(\ref{sepvariable}) can be therefore written in the Laplace domain as
\begin{equation}
U\left(  x,s\right)  =\sum_{n=1}^{+\infty}\phi_{n}\left(  x\right)
\frac{F_{n}\left(  x,s\right)  }{\mu s^{2}+J_{xz}\beta_{n}^{4}E\left(
s\right)  } \label{timesolution}%
\end{equation}

In particular, by considering as external applied force, a Dirac Delta of
constant amplitude $F_{0}$, in both the time and the spatial domains, the
force can be written as $f\left(  x,t\right)  =F_{0}\delta\left(
x-x_{f}\right)  \delta\left(  t-t_{0}\right)  $. Therefore in the Laplace
domain it becomes%
\begin{equation}
F_{n}=\int_{0}^{L}F_{0}\delta\left(  x-x_{f}\right)  \phi_{n}\left(  x\right)
\mathrm{d}x=F_{0}\phi_{n}\left(  x_{f}\right)  \label{Fn}%
\end{equation}
and finally the system response is%
\begin{equation}
U\left(  x,s\right)  =F_{0}\sum_{n=1}^{+\infty}\frac{\phi_{n}\left(  x\right)
\phi_{n}\left(  x_{f}\right)  }{\mu s^{2}+J_{xz}\beta_{n}^{4}E\left(
s\right)  } \label{system_response}%
\end{equation}
.

\section{Viscoelastic model - System eigenvalues}

Let us first consider an ideal viscoelastic material with a single relaxation
time $\tau_{1}$, whose elastic properties can be represented by the modulus%
\begin{equation}
E\left(  s\right)  =E_{0}+E_{1}\frac{\tau_{1}s}{1+\tau_{1}s}%
\label{modulus_1tau}%
\end{equation}
By substituting the previous complex function in Eq.(\ref{beta_solutions}),
the characteristic equation for each $n_{th}$ mode can be obtained
\begin{equation}
\tau_{1}s^{3}+s^{2}+\left(  E_{0}+E_{1}\right)  \tau_{1}r_{n}s+r_{n}%
E_{0}=0\label{characteristic equation}%
\end{equation}
where $r_{n}=\left(  \beta_{n}\right)  ^{4}J_{xz}/\mu$. Notice that the
solutions of the cubic equation Eq.(\ref{characteristic equation}) can be i)
one real root and two complex conjugate roots, ii) all roots real. This means
that one eigenvalue is always related to an overdamped motion. When the other
two eigenvalues are complex conjugate, they represent the oscillatory
contribute of the $n_{th}$ mode in the beam dynamics. Otherwise, in case of
three real roots, the $n_{th}$ mode is not oscillatory.\begin{figure}[ptb]
\begin{center}
\includegraphics[
height=6cm,
]{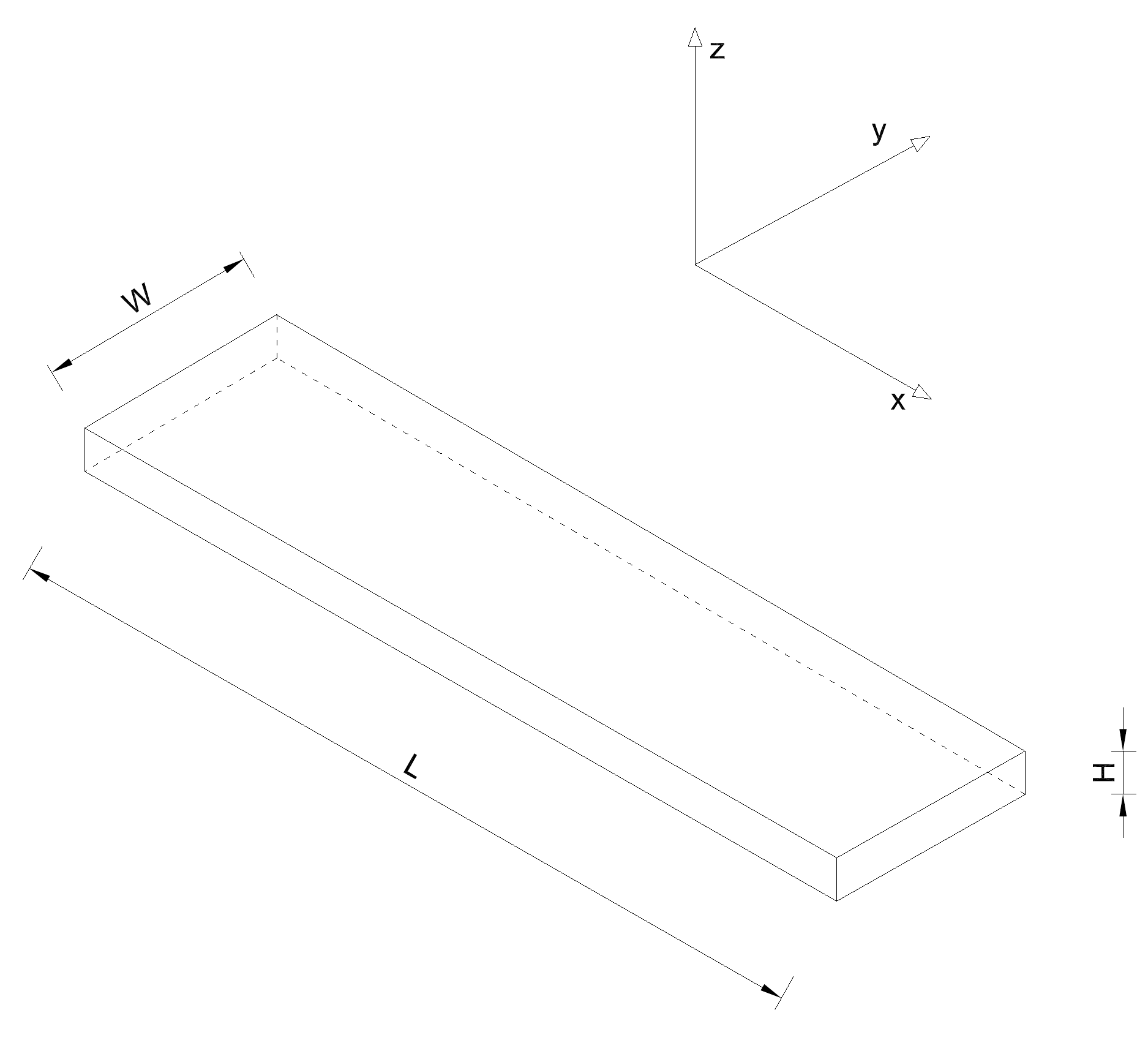}
\end{center}
\caption{Viscoelastic beam of lenght $L$, cross section area $A=WH$.}%
\label{Figure2}%
\end{figure}

With regards to the transverse motions of a narrow, homogenous beam with a
bending stiffness $E_{0}J_{xz}$ and density $\rho$, the value of the natural
frequencies can be calculated using a simple formula which is always valid,
regardless of the boundary conditions \cite{Thomson}:%
\begin{equation}
\omega_{n}=\left(  \frac{c_{n}}{L}\right)  ^{2}\sqrt{\frac{E_{0}J_{xz}}{\rho
A}} \label{general_freq}%
\end{equation}
where coefficient $c_{n}$ depends on the specific boundary conditions. The
first natural frequency, in particular, can be written as%
\begin{equation}
\omega_{1}=\alpha^{2}\delta_{1} \label{first_freq}%
\end{equation}
being $\delta_{1}=c_{1}^{2}\sqrt{E_{0}A/\left(  \rho J_{xz}\right)  }$, and
$\alpha=R_{g}/L$ the dimensionless beam length, with $R_{g}=\sqrt{J_{xz}/A}$
the radius of gyration. For the rectangular beam cross section under
investigation (Figure \ref{Figure2}), one has $\alpha=H/\left(  \sqrt
{12}L\right)  $ and $\delta_{1}=\left(  c_{1}^{2}/H\right)  \sqrt{12E_{0}%
/\rho}$ . The non-dimensional eigenvalue is now defined%
\begin{equation}
\bar{s}=s/\delta_{1} \label{nondim_eigenv}%
\end{equation}
and in particular one has, for the $n_{th}$ mode, $\omega_{n}^{2}=E_{0}%
\beta_{n}^{4}J_{xz}/\mu=r_{n}E_{0}$ and $\delta_{n}=c_{n}^{2}\sqrt
{E_{0}A/\left(  \rho J_{xz}\right)  }$.

By substituting Eq.(\ref{nondim_eigenv}) in Eq.(\ref{characteristic equation}%
), the following non-dimensional characteristic equation is obtained%
\begin{equation}
\bar{s}^{3}+\bar{s}^{2}\frac{1}{\theta_{1}}+\left(  1+\gamma_{1}\right)
\alpha^{4}\Delta_{n}^{2}\bar{s}+\frac{1}{\theta_{1}}\alpha^{4}\Delta_{n}^{2}=0
\label{nondim_charact_eq}%
\end{equation}
where $\Delta_{n}=\delta_{n}/\delta_{1}$, and having defined the dimensional
groups%
\begin{equation}
\theta_{1}=\delta_{1}\tau_{1} \label{teta}%
\end{equation}%
\begin{equation}
\gamma_{1}=E_{1}/E_{0} \label{gamma}%
\end{equation}
. Eq.(\ref{nondim_charact_eq}) can be then re-written as
\begin{equation}
\bar{s}^{3}+\sum_{j=0}^{2}a_{j}^{j}\bar{s}^{j}=0 \label{nondim_chareq_compact}%
\end{equation}
where $a_{0}=\left(  1/\theta_{1}\right)  \alpha^{4}\Delta_{n}^{2}$,
$a_{1}=\left(  1+\gamma_{1}\right)  \alpha^{4}\Delta_{n}^{2}$, $a_{2}%
=1/\theta_{1}$.

By defining%
\begin{equation}
Q=\frac{3a_{1}-a_{2}^{2}}{9} \label{Qval}%
\end{equation}%
\begin{equation}
R=\frac{9a_{2}a_{1}-27a_{0}-2a_{2}^{3}}{54} \label{Rval}%
\end{equation}
the discriminant of Eq.(\ref{nondim_chareq_compact}) is $D=Q^{3}+R^{2}$, and
the solutions of Eq.(\ref{nondim_chareq_compact}) can be therefore written as
\cite{Abramowitz1965}%
\begin{align}
\bar{s}_{1}  &  =-\frac{a_{2}}{3}-\frac{1}{2}\left(  S+T\right)
+\mathrm{i}\frac{\sqrt{3}}{2}\left(  S-T\right) \label{SolutionsEqChar}\\
\bar{s}_{2}  &  =-\frac{a_{2}}{3}-\frac{1}{2}\left(  S+T\right)
-\mathrm{i}\frac{\sqrt{3}}{2}\left(  S-T\right) \nonumber\\
\bar{s}_{3}  &  =-\frac{a_{2}}{3}+\left(  S+T\right) \nonumber
\end{align}
where $S=\sqrt[3]{R+\sqrt{D}}$ and $T=\sqrt[3]{R-\sqrt{D}}$. In our case, the
discriminant $D$, indicated as $D_{1}\left(  n\right)  $, is function of $n$,
i.e. of the number of the $n_{th}$ mode considered
\begin{equation}
D_{1}\left(  n\right)  =\frac{\alpha^{4}\Delta_{n}^{2}\left\{  4+\alpha
^{4}\Delta_{n}^{2}\theta_{1}^{2}\left[  8-\gamma_{1}\left(  20+\gamma
_{1}\right)  +4\alpha^{2}\left(  1+\gamma_{1}\right)  ^{3}\Delta_{n}^{2}%
\theta_{1}^{2}\right]  \right\}  }{108\theta_{1}^{4}} \label{discriminant_1tr}%
\end{equation}
This function $D_{1}\left(  n\right)  $ plays a key role in the understanding
the nature of the roots of Eq.(\ref{nondim_chareq_compact}), as it will be
widely discussed in Section III.

At last, the beam cross-section acceleration $A\left(  x,s\right)  $ in terms
of the above defined non-dimensional groups is formulated, considering that
$A\left(  x,s\right)  =s^{2}U\left(  x,s\right)  $ (see
Eq.(\ref{system_response}))%

\begin{equation}
A\left(  x,\bar{s}\right)  =F_{0}\sum_{n=1}^{+\infty}\frac{\bar{s}^{2}\left(
1+\theta_{1}\bar{s}\right)  \phi_{n}\left(  x\right)  \phi_{n}\left(
x_{f}\right)  }{\mu\theta_{1}\left(  \bar{s}^{3}+\sum_{j=0}^{2}a_{j}\bar
{s}^{j}\right)  }\label{SystResp_adim}%
\end{equation}
being $\phi_{n}\left(  x\right)  $ the eigenfunctions defined in
Eq.(\ref{modes}).\begin{figure}[ptb]
\begin{center}
\includegraphics[
height=7cm,
]{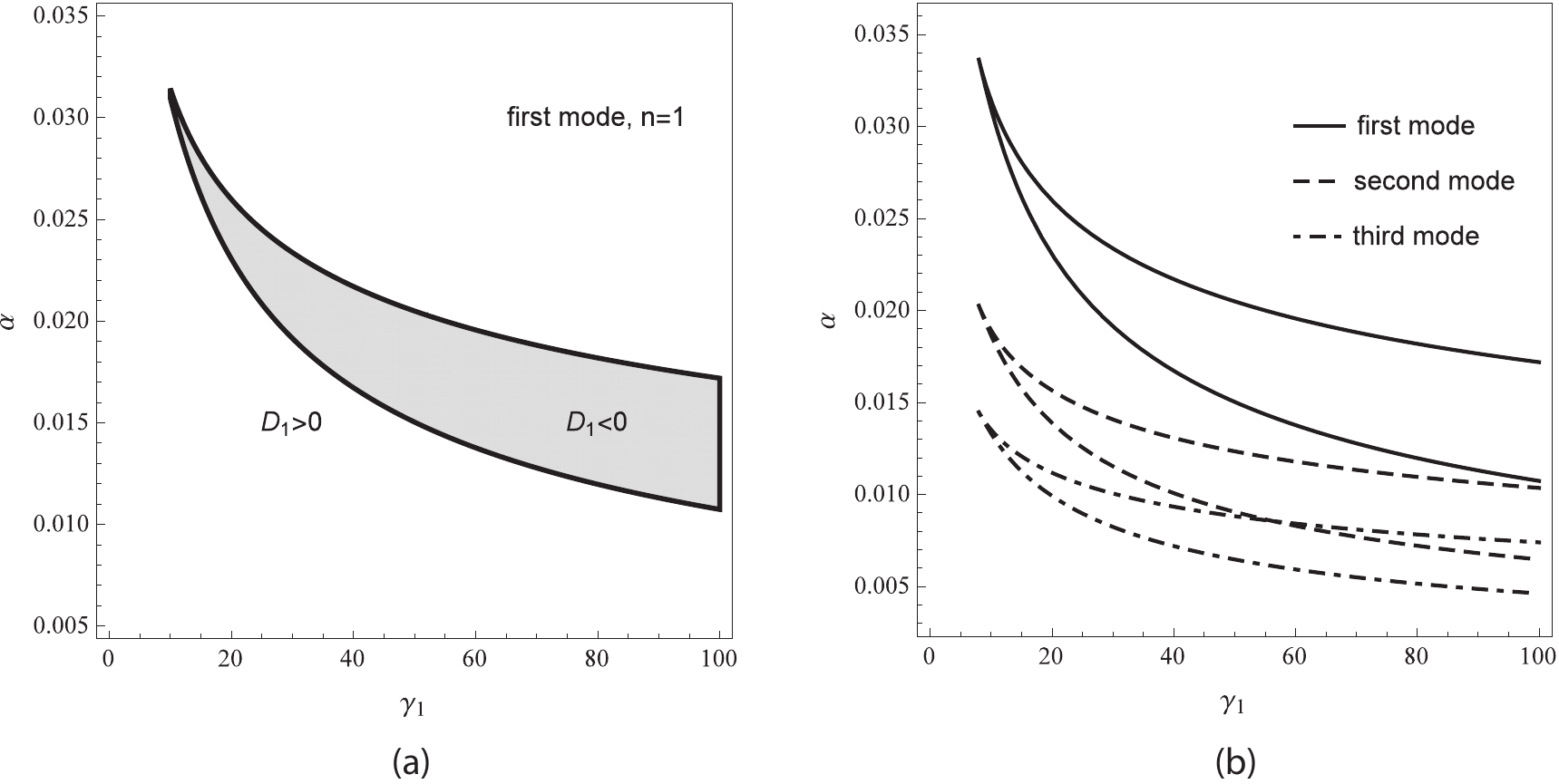}
\end{center}
\caption{The region map for the first flexural mode of the beam $n=1$, for
$\theta_{1}=\bar{\theta}_{1}$ (a). The shaded area indicates the parameter
$\left(  \alpha,\gamma_{1}\right)  $ combinations which determine the
suppression of the first peak. \ Similar region maps are shown, for the first
three modes $n=1,2,3$ (b). Interestingly, for several values of $\alpha$ and
$\gamma_{1}$, some areas at $D_{1}\left(  n\right)  <0$ are overlapped,
determining the suppression of more peaks simultaneously.}%
\label{Figure3}%
\end{figure}

More realistically, a second relaxation time contribute is now included in the
viscoelastic modulus $E\left(  s\right)  $, which therefore becomes%
\begin{equation}
E\left(  s\right)  =E_{0}+E_{1}\frac{\tau_{1}s}{1+\tau_{1}s}+E_{2}\frac
{\tau_{2}s}{1+\tau_{2}s} \label{modulus_2tau}%
\end{equation}
Following the same approach previously described, the fourth-order
characteristic equation for each $n_{th}$ mode can be obtained
\begin{equation}
\bar{s}^{4}+\sum_{j=0}^{3}a_{j}\bar{s}^{j}=0 \label{nondim_chareq_comp_2tr}%
\end{equation}
where%
\begin{align}
a_{0}  &  =\alpha^{4}\Delta_{n}^{2}\frac{1}{\theta_{1}\theta_{2}%
}\label{coeff_4orderEq}\\
a_{1}  &  =\left(  \frac{1}{\theta_{2}}+\frac{1}{\theta_{1}}+\frac{1}%
{\theta_{2}}\gamma_{1}+\frac{1}{\theta_{1}}\gamma_{2}\right)  \alpha^{4}%
\Delta_{n}^{2}\nonumber\\
a_{2}  &  =\left(  \frac{1}{\theta_{1}\theta_{2}}+\alpha^{4}\Delta_{n}%
^{2}+\alpha^{4}\Delta_{n}^{2}\gamma_{1}+\alpha^{4}\Delta_{n}^{2}\gamma
_{2}\right) \nonumber\\
a_{3}  &  =\left(  \frac{1}{\theta_{1}}+\frac{1}{\theta_{2}}\right) \nonumber
\end{align}
having defined $\gamma_{2}=E_{2}/E_{0}$ and $\theta_{2}=\tau_{2}\delta_{1}$.
Moreover, it is possible to define, for the quartic equation
Eq.(\ref{nondim_chareq_comp_2tr}), the discriminant $D_{2}\left(  n\right)  $
\cite{Lazard1988}-\cite{Rees1922}%
\begin{align}
D_{2}\left(  n\right)   &  =256a_{0}^{3}-192a_{3}a_{1}a_{0}^{2}-128a_{2}%
^{2}a_{0}^{2}+144a_{2}a_{1}^{2}a_{0}-27a_{1}^{4}+144a_{3}^{2}a_{2}a_{0}%
^{2}-6a_{3}^{2}a_{1}^{2}a_{0}-80a_{3}a_{2}^{2}a_{1}a_{0}%
+\label{discriminant_2tr}\\
&  +18a_{3}a_{2}a_{1}^{3}+16a_{2}^{4}a_{0}-4a_{2}^{3}a_{1}^{2}-27a_{3}%
^{4}a_{0}^{2}+18a_{3}^{3}a_{2}a_{1}a_{0}-4a_{3}^{3}a_{1}^{3}-4a_{3}^{2}%
a_{2}^{3}a_{0}+a_{3}^{2}a_{2}^{2}a_{1}^{2}\nonumber
\end{align}
which can be utilized to deduce important properties of the roots of
Eq.(\ref{nondim_chareq_comp_2tr}).

The beam cross-section acceleration $A\left(  x,\bar{s}\right)  $ is in this
case%
\begin{equation}
A\left(  x,\bar{s}\right)  =F_{0}\sum_{n=1}^{+\infty}\frac{\bar{s}^{2}\left(
1+\theta_{1}\bar{s}\right)  \left(  1+\theta_{2}\bar{s}\right)  \phi
_{n}\left(  x\right)  \phi_{n}\left(  x_{f}\right)  }{\mu\theta_{1}\theta
_{2}\left(  \bar{s}^{4}+\sum_{j=0}^{3}a_{j}\bar{s}^{j}\right)  }%
\label{SystResp_adim_2tr}%
\end{equation}
\begin{figure}[ptb]
\begin{center}
\includegraphics[
height=6cm,
]{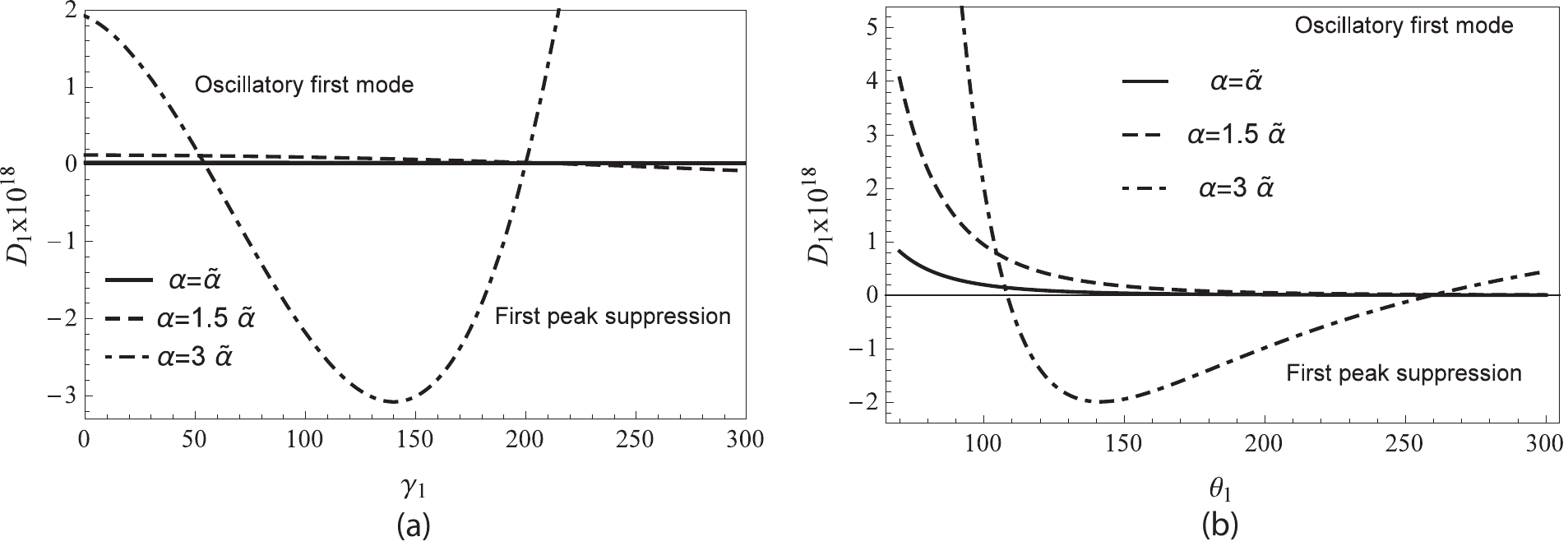}
\end{center}
\caption{The discriminant $D_{1}\left(  1\right)  $ for the first mode $n=1$,
as a function of $\gamma_{1}$, for $\theta_{1}=\bar{\theta}_{1}$ (a), and as a
function of $\theta_{1}$, for $\gamma_{1}=\bar{\gamma}_{1}$ (b), for different
values of $\alpha$, i.e $\alpha=\tilde{\alpha}$ (solid line), $\alpha
=1.5~\tilde{\alpha}$ (dashed line), $\alpha=3~\tilde{\alpha}$ (dot-dashed
line). }%
\label{Figure4}%
\end{figure}

\section{Results}

In this section the main results of the presented analysis are discussed. The
flexural vibrations of a viscoelastic beam with rectangular cross section and
thickness $H=1~[\mathrm{cm}]$, which oscillates in the $xz$-plane (Figure
\ref{Figure2}) are studied. The only geometrical parameter which is considered
varying in calculations, is the beam length $L$. In particular, the ratio
$\alpha=R_{g}/L$ is changed maintaining $R_{g}=H/\sqrt{12}$ constant. The main
scope of the paper is not a quantitative investigation of a specific
viscoelastic material, but a qualitative study of a generic viscoelastic beam
behaviour, which can be considered at different lengths $L$ (e.g. in
experimental testing campaigns, to cover wide frequency ranges) and at
different working temperatures (i.e. with varying elastic coefficients $E_{k}$
and relaxation times $\tau_{k}$). In this view, the two material properties
considered constant in the numerical calculations are $\rho
=1180~[\mathrm{kg\ m}^{-3}]$ and $E_{0}=2.24\ast10^{6}~[\mathrm{Pa}]$ of a
typical viscoelastic material, i.e. PMMA (polymethyl methacrylate)
\cite{Schapery}. Therefore the parameters $\delta_{n}$ are constant and, in
particular, $\delta_{1}=3.4\ast10^{5}$ for the first flexural mode of the
beam. The other properties $E_{1}$, $E_{2}$, $\tau_{1}$ and $\tau_{2}$ are
taken varying in the analysis, however $\bar{\gamma}_{1}=E_{1}/E_{0}=87$ and
$\bar{\gamma}_{2}=E_{2}/E_{0}=126~$of PMMA are considered as reference.
Moreover, the relaxation times for the frequency range under study give the
reference values $\bar{\theta}_{1}=\delta_{1}\tau_{1}=170$, $\bar{\theta}%
_{2}=\delta_{1}\tau_{2}=6700$. The numerical values here considered, are
simply representative of a real viscoelastic material but, thanks to the
dimensional analysis presented in the paper, can be substituted with the
constants of any other viscoelastic material, thus not modifying the
qualitative results.

At first, let us consider an ideal viscoelastic material with one relaxation
time $\tau_{1}$, and elastic coefficients $E_{0}$ and $E_{1}$
(Eq.\ref{modulus_1tau}). For each $n_{th}$ mode, the three eigenvalues (see
Eq.(\ref{SolutionsEqChar})) can be calculated. The two complex conjugate
eigenvalues represent the oscillatory counterpart of the beam $n_{th}$ mode.
\ The real eigenvalue gives rise to a pure dissipative contribute. However,
when the discriminant $D_{1}\left(  n\right)  $, defined in
Eq.(\ref{discriminant_1tr}), is negative $D_{1}\left(  n\right)  <0$, all
roots of Eq.(\ref{nondim_charact_eq}) are real, and the $n_{th}$ mode is not
oscillatory. In Figure \ref{Figure3}-a, a region map is shown, for the first
flexural mode of the beam ($n=1$), with $\theta_{1}=\bar{\theta}_{1}$.
\begin{figure}[ptb]
\begin{center}
\includegraphics[
height=6cm,
]{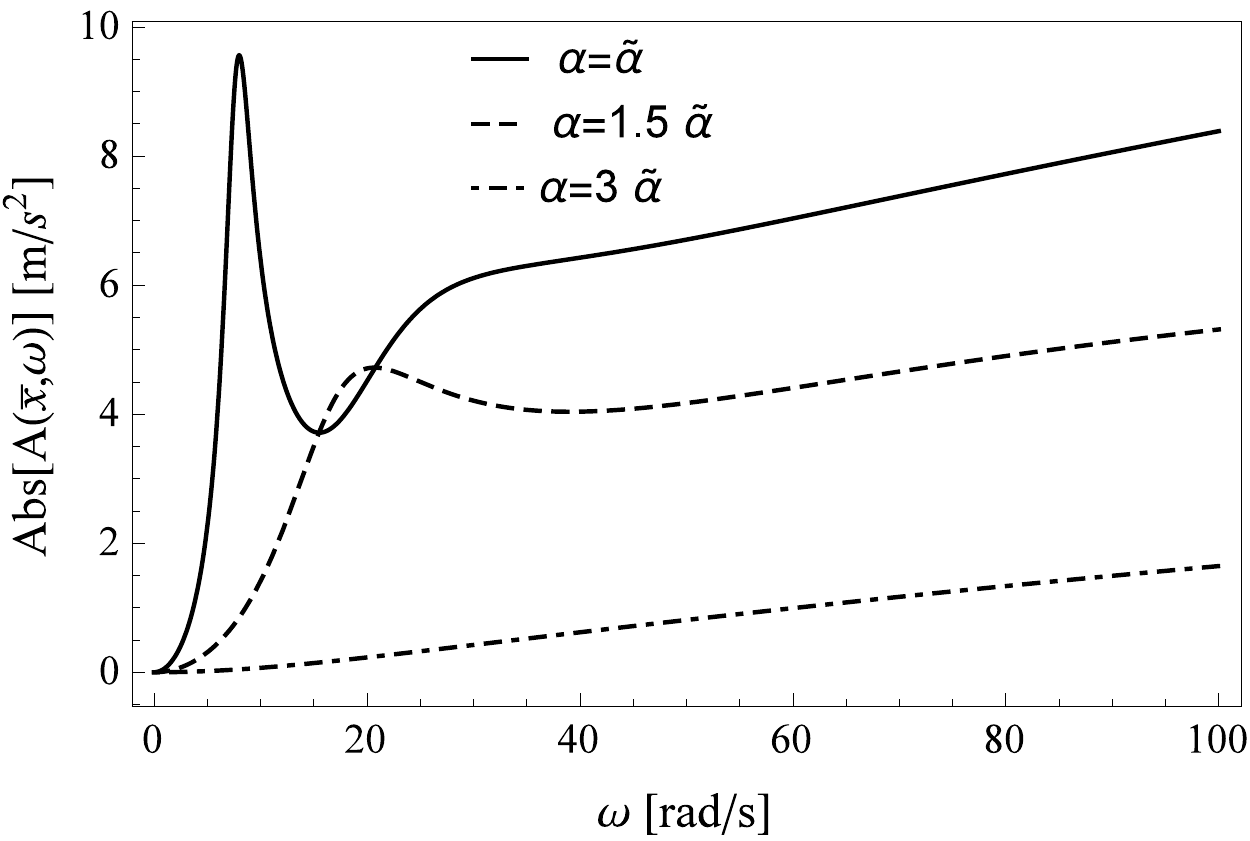}
\end{center}
\caption{The acceleration modulus $\left\vert A\left(  \bar{x},\omega\right)
\right\vert $ of the viscoelastic beam with one relaxation time, in the
section $x=x_{f}=\bar{x}=0.4L$, for $\theta_{1}=\bar{\theta}_{1}$, $\gamma
_{1}=\bar{\gamma}_{1}$, and for three different values of $\alpha$, i.e.
$\alpha=\tilde{\alpha}$ (solid line), $\alpha=1.5\tilde{\alpha}$ (dashed
line), $\alpha=3\tilde{\alpha}$ (dot-dashed line). For $\alpha=3\tilde{\alpha
}$, being $D_{1}\left(  1\right)  <0$ , the first peak is suppressed.}%
\label{Figure5}%
\end{figure}The shaded area is obtained with the parameter values $\left(
\alpha,\gamma_{1}\right)  $ which give the condition $D_{1}\left(  1\right)
<0$. Analogous maps, for the first three flexural modes of the beam, are
represented by the correspondent curves in Figure \ref{Figure3}-b, which are
obtained by finding the two real solutions $\alpha=\alpha\left(  \gamma
_{1}\right)  $ of the equation $D_{1}\left(  n\right)  =0$, for $n=1,2,3$. By
properly combining the parameters $\left(  \alpha,\gamma_{1}\right)  $,
different peaks can be suppressed simultaneously, since the areas which give
the condition $D_{1}\left(  n\right)  <0$ for different values of $n$ are
overlapped. In particular, this means that, once the material is prescribed,
i.e. for given values of $\theta_{1}$ and $\gamma_{1}$, the dynamics of the
beam can be decisively modified by varying its length $L$. The sign of the
discriminant $D_{1}\left(  1\right)  $, for the first mode, can be directly
deduced by means of the curves plotted in Figure \ref{Figure4}, where
$D_{1}\left(  1\right)  $ is shown as a function of $\gamma_{1}$ (Figure
\ref{Figure4}-a), for $\theta_{1}=\bar{\theta}_{1}$, and as a function of
$\theta_{1}$, for $\gamma_{1}=\bar{\gamma}_{1}$ (Figure \ref{Figure4}-b), for
different values of $\alpha$, i.e. $\alpha=\tilde{\alpha}$ (solid line),
$\alpha=1.5\tilde{\alpha}$ (dashed line), $\alpha=3\tilde{\alpha}$ (dot-dashed
line), where it has been considered the reference beam length equal to
$\tilde{L}=60[\mathrm{cm}]$, and therefore $\tilde{\alpha}=R_{g}/\tilde
{L}=0.0048$. In Figure \ref{Figure5} the system response is represented, in
terms of the acceleration modulus $\left\vert A\left(  \bar{x},\omega\right)
\right\vert $ (see Eq.\ref{SystResp_adim}), evaluated at the beam section
$x=x_{f}=\bar{x}=0.4L$, for $\theta_{1}=\bar{\theta}_{1}$ and $\gamma_{1}%
=\bar{\gamma}_{1}$. Three different values of beam length $L$ are considered,
i.e. $\alpha=\tilde{\alpha}$ (solid line), $\alpha=1.5\tilde{\alpha}$ (dashed
line), $\alpha=3\tilde{\alpha}$ (dot-dashed line), which give a clear first
peak for $\alpha=\tilde{\alpha}$ and $\alpha=1.5\tilde{\alpha}$, a suppressed
first peak for $\alpha=3\tilde{\alpha}$, being $D_{1}\left(  1\right)  <0$ in
the last case, as one can notice in Figure \ref{Figure4}.\begin{figure}[ptb]
\begin{center}
\includegraphics[
height=7cm,
]{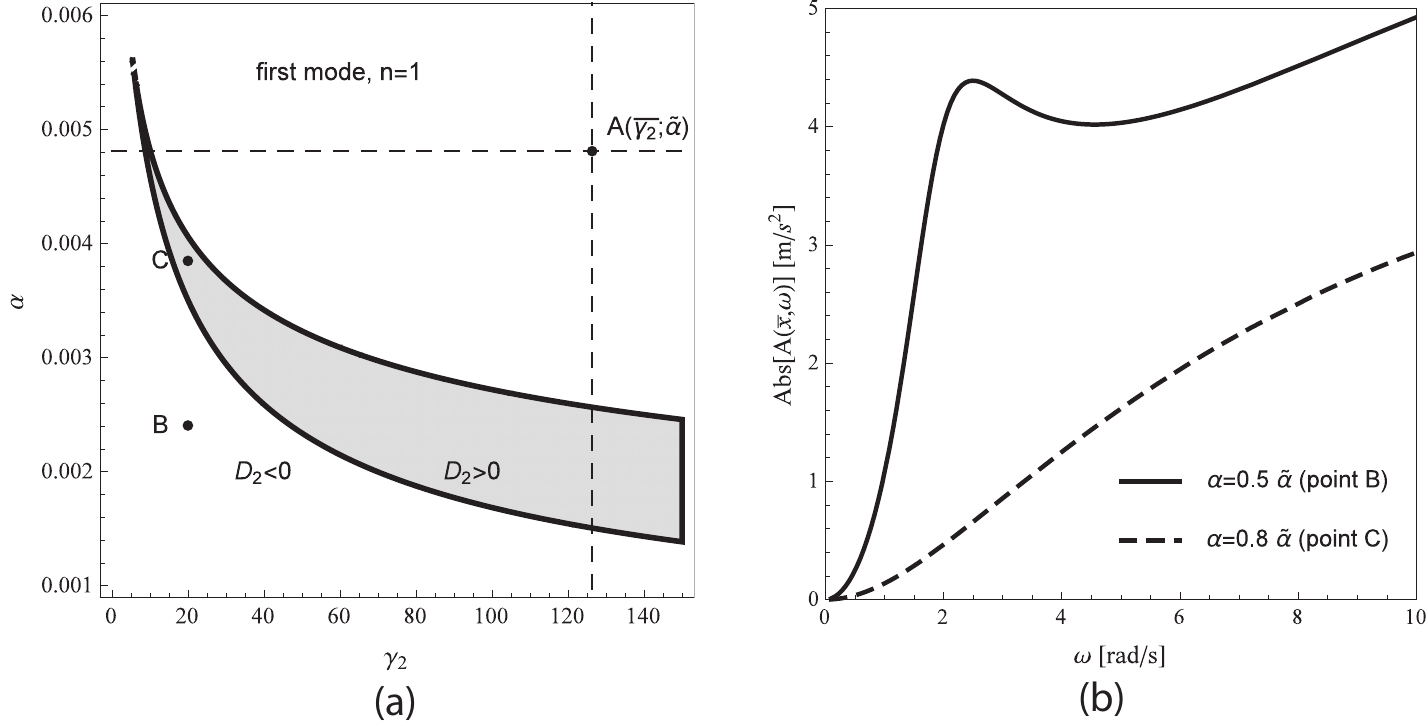}
\end{center}
\caption{The region map for the first flexural mode of the beam $n=1$, with
two relaxation times, for $\theta_{1}=$ $\bar{\theta}_{1}$, $\theta_{2}=$
$\bar{\theta}_{2}$ , $\gamma_{1}=$ $\bar{\gamma}_{1}$: the shaded area
indicates the parameter $\left(  \gamma_{2},\alpha\right)  $ combinations
which determine the suppression of the first peak (a). The acceleration
modulus $\left\vert A\left(  \bar{x},\omega\right)  \right\vert $ of the
viscoelastic beam with two relaxation times is plotted, in the beam section
$x=x_{f}=\bar{x}=0.4L$, related to points $B(20;0.5\tilde{\alpha})$ and
$C(20;0.8\tilde{\alpha})$ .}%
\label{Figure6}%
\end{figure}

Let us consider now a viscoelastic material with two relaxation times, in
order to explore possible variations in the beam dynamics. In Figure
\ref{Figure6}-a a region map is shown, for the first flexural mode of the beam
($n=1$), with $\theta_{1}=\bar{\theta}_{1}$, $\theta_{2}=$ $\bar{\theta}_{2}$
, $\gamma_{1}=$ $\bar{\gamma}_{1}$. The shaded area is obtained with the
parameter values $\left(  \gamma_{2},\alpha\right)  $ compounds which give the
condition $D_{2}\left(  1\right)  >0$. Indeed, since the nature of the
eigenvalues characteristic equation is changed, i.e. being
Eq.(\ref{nondim_chareq_comp_2tr}) a quartic equation, the condition for an
oscillatory mode is $D_{2}\left(  n\right)  <0$ \cite{Lazard1988}%
-\cite{Rees1922}. Point $A$ in Figure \ref{Figure6}-a is obtained by
considering the reference parameters $\left(  \bar{\gamma}_{2},\tilde{\alpha
}\right)  $. Since it is far from the shaded area at $D_{2}\left(  1\right)
>0$, it is related to a oscillatory first mode for small $\alpha$\ variations.
In order to better evaluate the influence of the parameter $\alpha$, points
$B$ and $C$ have been considered to calculate the frequency response of the
beam. In Figure \ref{Figure6}-b, the acceleration modulus $\left\vert A\left(
\bar{x},\omega\right)  \right\vert $ (see Eq.\ref{SystResp_adim_2tr}),
calculated at the beam section $x=x_{f}=\bar{x}=0.4L$, is shown for
$\theta_{1}=\bar{\theta}_{1}$, $\theta_{2}=$ $\bar{\theta}_{2}$ , $\gamma
_{1}=$ $\bar{\gamma}_{1}$, $\gamma_{2}=20$ and for $\alpha=0.5\tilde{\alpha}$
(solid line, point $B$ of Figure \ref{Figure6}-a), $\alpha=0.8\tilde{\alpha}$
(dashed line, point $C$ of Figure \ref{Figure6}-a). It is possible to observe
that the curve obtained with $\alpha=0.8\tilde{\alpha}$ does not present the
first peak, because of the considered parameters, which give in this case
$D_{2}\left(  1\right)  >0$.\begin{figure}[ptb]
\begin{center}
\includegraphics[
height=6cm,
]{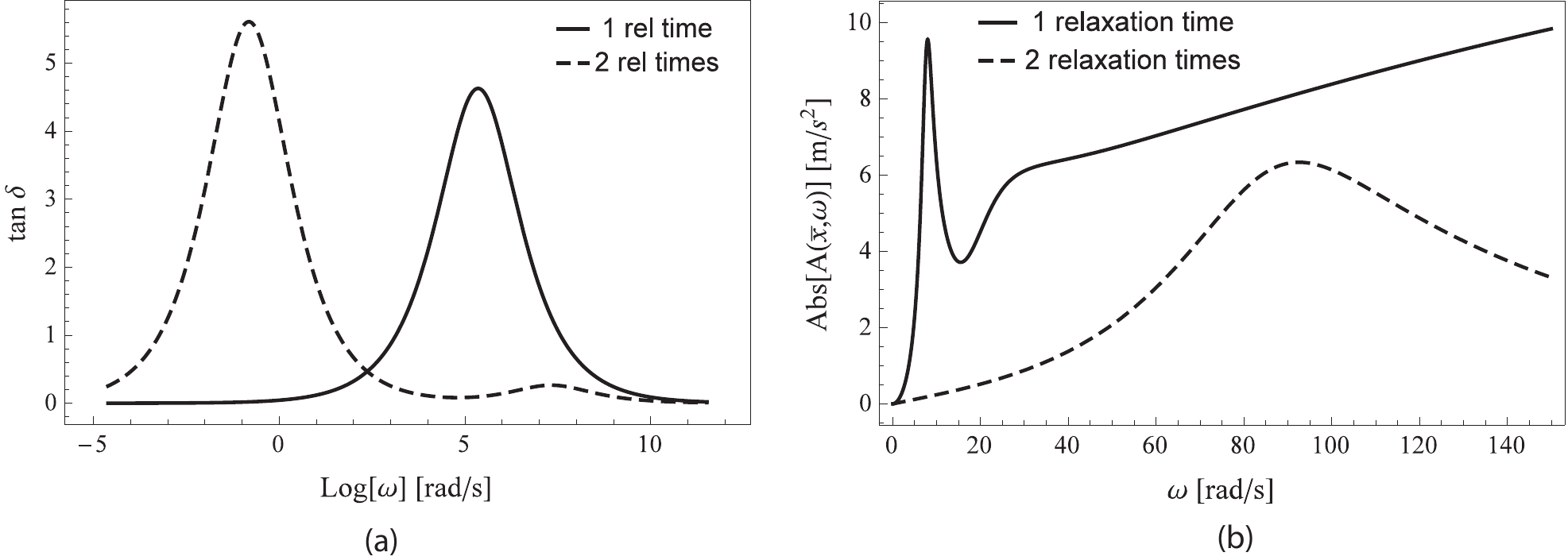}
\end{center}
\caption{The viscoelastic modulus $E(\omega)$, for one relaxation time, with
$\theta_{1}=\bar{\theta}_{1}$, $\gamma_{1}=$ $\bar{\gamma}_{1}$ (solid curve),
and for two relaxation times, with $\theta_{1}=\bar{\theta}_{1}$, $\gamma
_{1}=$ $\bar{\gamma}_{1}$, $\theta_{2}=\bar{\theta}_{2}$, $\gamma_{2}=$
$\bar{\gamma}_{2}$ (dashed curve), represented by the function $\tan
\delta=\operatorname{Re}[E(\omega)]/\operatorname{Im}[E(\omega)]$ (a); the
acceleration modulus $\left\vert A\left(  \bar{x},\omega\right)  \right\vert $
is shown in the two cases, for $\alpha=\tilde{\alpha}$, $x=x_{f}=\bar{x}%
=0.4L$, for one relaxation time (solid curve) and for two relaxation times
(dashed curve) (b).}%
\label{Figure7}%
\end{figure}

\section{Frequency responses}

Let us observe that, besides the sign of the discriminants $D_{1}\left(
n\right)  $ and $D_{2}\left(  n\right)  $, and the different values of
$\alpha$, which establish the possible $n_{th}$ peak suppression, there is no
significant difference in considering one or two relaxation times. In this
respect, the spectrum of the viscoelastic modulus $E(\omega)$ is considered in
the two cases, as shown in Figure \ref{Figure7}-a, where the function
$\tan\delta=\operatorname{Re}[E(\omega)]/\operatorname{Im}[E(\omega)]$ is
plotted, for one relaxation time ($\theta_{1}=\bar{\theta}_{1}$, $\gamma_{1}=$
$\bar{\gamma}_{1}$, solid curve), and for two relaxation times ($\theta
_{1}=\bar{\theta}_{1}$, $\gamma_{1}=$ $\bar{\gamma}_{1}$, $\theta_{2}%
=\bar{\theta}_{2}$, $\gamma_{2}=$ $\bar{\gamma}_{2}$, dashed curve). It is
evident that in the two cases, the transition region, where the function
$\tan\delta$ reaches the maximum, thus determining the prominent energy
dissipation, is differently positioned in the frequency spectrum. The
correspondent acceleration moduli $\left\vert A\left(  \bar{x},\omega\right)
\right\vert $ are shown in Figure \ref{Figure7}-b, for $\alpha=\tilde{\alpha}%
$, $x=x_{f}=\bar{x}=0.4L$, for one relaxation time (solid curve) and for two
relaxation times (dashed curve). Notice that, in the case of two relaxation
times the function $\tan\delta$ presents higher values, with respect to the
one relaxation time case, in the range of frequencies where the first peak
lies ($0-100~[\mathrm{rad\ s}^{-1}]$). This is why the first peak, when two
relaxation times are\ considered, is more damped. It is important to underline
that, because of the intrinsic characteristics of viscoelastic materials
\cite{Christensen}, which see the viscoelastic modulus $E(\omega)$ depending
on temperature, the above mentioned damping effect can be observed just
modifying the surrounding temperature. Indeed, increasing or decreasing the
working temperature, the functions $\operatorname{Re}[E(\omega)]$ and
$\operatorname{Im}[E(\omega)]$ are shifted towards higher or smaller
frequencies respectively (as well as and the function $\tan\delta)$, and
consequently the material damping is differently spread in frequency.

However, once the material is prescribed, i.e. the viscoelastic modulus
$E(\omega)$ is defined with the related parameters, $\theta_{1}$ and
$\gamma_{1}$ for one relaxation time, $\theta_{1}$, $\theta_{2}$, $\gamma_{1}$
and $\gamma_{2}$ for two relaxation times, the dimensionless beam length
$\alpha$ plays a crucial role in the possible overlapping of the first natural
frequency $\omega_{1}$ with the transition region. Moreover, in the hypothesis
of linearity, such considerations can be extended to all the peaks, since the
system can be decoupled and each vibration mode can be studied independently.

In conclusion, through the defined dimensionless parameters, it is possible to
completely disclose the transversal vibrations of a viscoelastic beam.
Suppressing certain peaks, by varying the beam length with $\alpha$, or by
changing the material properties (i.e. $\theta_{1}$ and $\gamma_{1}$ for one
relaxation time, $\theta_{1}$, $\theta_{2}$, $\gamma_{1}$ and $\gamma_{2}$ for
two relaxation times) for example by modifying the surrounding temperature, is
an appealing chance for different applications. In particular, it is important
to stress that, although the real viscoelastic materials present more than two
relaxation times, the number of relaxation times to be considered in modelling
the beam dynamics, does not represent a limit of our study. Firstly, because
it has been shown that there is not a considerable difference, from a
qualitative point of view, by increasing the number of time relaxations.
Furthermore, it is always possible to divide the frequency spectrum under
analysis in several intervals, thus allowing the decreasing of the
predominant$\ $time relaxations number in such intervals. Moreover, by varying
the beam length, it is possible to study a wide frequency range, by focusing
the attention only to the first resonant peaks, so that the (Euler-Bernoulli)
hypothesis still remains valid.

Finally, it must be pointed out that this study can be utilized to properly
interpret the viscoelastic beam vibrational spectrum, when a material
characterization is carried out. This is an awkward task, indeed, since when a
viscoelastic beam with an unknown material is experimentally studied, the
resonant peaks positions are not so straightforward to be identified, as in
the elastic case. This kind of experimental investigation is currently object
of study, with the aim to characterize viscoelastic materials by means of the
transversal vibrations of beams with different lengths.

\section{Conclusions}

In this paper an analytical study of the transversal vibrations of a
viscoelastic beam has been presented. The analytical solution has been
obtained by means of modal superposition. In particular, while the beam
eigenfunctions are the same of the perfectly elastic case, the eigenvalues
strongly depend on the material viscoelasticity, and they increase in number
with the relaxation times of the viscoelastic modulus. In order to put in
evidence the main characteristics of the beam dynamics, two cases have been
considered, i.e. a viscoelastic material both with one single relaxation time
and with two relaxation times. A dimensional analysis has been performed,
which has disclosed the fundamental parameters involved in the vibrational
behaviour of the beam. Such parameters depend on both the material properties
and the beam geometry. Some new characteristic maps related to the eigenvalues
nature of the studied system have been provided, that can be drawn for each
natural frequency of the beam. In comparison to the existing maps presented in
literature for a sdof system, these maps may help in determining the parameter
compounds needed to enhance or suppress certain frequency peaks, one at a time
or more simultaneously, and the same approach can be exploited for any kind of
mdof system. Interestingly, it has been observed that, by maintaining constant
the thickness of the beam cross section, the dimensionless beam length can be
utilized as key parameter to properly adjust the resonant peaks, once the
material has been selected. The presented study, hence, enables to
conveniently design a viscoelastic beam, in order to obtain the most suitable
dynamics in the frequency range of interest, thus becoming a powerful tool for
many applications, from system damping control to materials characterization.

\end{document}